\documentclass[aps,pra,10pt,twocolumn,showpacs]{revtex4-1}
\usepackage[T1]{fontenc}
\usepackage{times}
\usepackage{amsfonts,amsmath,amssymb}
\usepackage{graphicx}
\usepackage[pdfstartview=FitH,colorlinks=true,linkcolor=blue,citecolor=blue,urlcolor=blue]{hyperref}

\begin{document}

\title{Optomechanical signature of a frictionless flow of superfluid light}

\author{Pierre-\'Elie Larr\'e}
\email{pierre.larre@unitn.it}
\author{Iacopo Carusotto}
\email{carusott@science.unitn.it}
\affiliation{INO-CNR BEC Center and Dipartimento di Fisica, Universit\`a di Trento, Via Sommarive 14, I-38123 Povo, Italy}

\date{\today}

\begin{abstract}
We propose an experimental setup that should make it possible to reveal the frictionless flow of a superfluid of light from the suppression of the drag force that it exerts onto a material obstacle. In the paraxial-propagation geometry considered here, the photon-fluid dynamics is described by a wave equation analogous to the Gross--Pitaevskii equation of dilute Bose--Einstein condensates and the obstacle consists in a solid dielectric slab immersed into a nonlinear optical liquid. By means of an ab initio calculation of the electromagnetic force experienced by the obstacle, we anticipate that superfluidity is detectable in state-of-the-art experiments from the disappearance of the optomechanical deformation of the obstacle.
\end{abstract}

\pacs{42.65.-k, 
      42.65.Sf, 
      42.50.Wk, 
      47.37.+q} 

\maketitle

\section{Introduction}
\label{Sec:Introduction}

Superfluidity, the capability of a fluid to flow without friction along a pipe or past an obstacle \cite{Leggett1999}, is undoubtedly among the most striking phenomena occurring in low-temperature liquids or gases. Since its first discovery in \textsuperscript{4}He \cite{Kapitza1938, Allen1938}, it has been observed in several other systems such as \textsuperscript{3}He \cite{Osheroff1972} or bosonic and fermionic ultracold atomic vapors \cite{Dalfovo1999, Pitaevskii2003}.

Following pioneering theoretical works \cite{Pomeau1993, Staliunas1993, Vaupel1996, Chiao1999, Chiao2000, Bolda2001, Carusotto2004, Leboeuf2010, Carusotto2014}, superfluidity has been experimentally demonstrated \cite{Amo2009} also in the completely-different optical context of the so-called quantum fluids of light. In suitable optical devices, a many-photon light beam can in fact behave collectively as a quantum fluid \cite{Carusotto2013}: Effective photon-photon interactions are mediated by the Kerr optical nonlinearity of the underlying medium, while photon confinement in a microcavity configuration or diffraction in a paraxial-propagation geometry provide a mass to the photon.

A transparent way to probe the superfluid properties of the photon fluid is to introduce a spatially localized defect into its flow and look at the perturbation that this latter generates into the fluid. Depending on the relative value of the flow speed compared to the sound speed, a full crossover has been revealed from a low-velocity superfluid regime, in which the flow remains practically unaffected by the presence of the obstacle \cite{Amo2009}, to a large-velocity regime characterized by the Cherenkov emission of Bogoliubov-like linear waves in the fluid and/or by the hydrodynamic nucleation of nonlinear excitations such as quantized vortices \cite{Nardin2011, Sanvitto2011} or dark solitons \cite{Amo2011, Grosso2011}. 

While the drop of the drag force experienced by the obstacle is among the main signatures of superfluidity in material fluids \cite{Rayfield1966, Allum1977, Castelijns1986, Raman1999, Onofrio2000, Miller2007}, so far all experiments on quantum fluids of light have only focused on the density and current disturbances induced by the obstacle in the flowing photon fluid \cite{Amo2009, Carusotto2013, Nardin2011, Sanvitto2011, Amo2011, Grosso2011}. In the wake of works on the classical \cite{Pavloff2002, Astrakharchik2004} and quantum \cite{Roberts2005} drag force in material fluids, calculations of the drag force in fluids of light were theoretically carried out by several authors \cite{Wouters2010, Berceanu2012, Larre2012, Larre2013} but no concrete experimental setup to effectively measure it has ever been proposed. The purpose of this work is to fill this gap and propose a configuration where the drag force generated by a flowing photon fluid onto an obstacle may be actually measured.

As compared to the planer-microcavity architecture used in the superfluid-light experiments of Refs.~\cite{Amo2009, Carusotto2013, Nardin2011, Sanvitto2011, Amo2011, Grosso2011}, the paraxial-propagation geometry, based on a bulk nonlinear optical medium and originally proposed in \cite{Pomeau1993}, appears most promising in view of this objective. We specifically consider the case of a monochromatic coherent electromagnetic wave propagates through a bulk Kerr nonlinear optical medium. Within the well-known \cite{Rosanov2002, Agrawal2007} reformulation of the paraxial propagation of light in terms of the Gross--Pitaevskii equation for the order parameter of a dilute Bose--Einstein condensate \cite{Dalfovo1999, Pitaevskii2003}, superfluidity is apparent as a suppression of scattering from regions characterized by spatial modulations of the refractive index \cite{Pomeau1993, Carusotto2014}. A first experiment to characterize the Bogoliubov dispersion of sound waves on top of a fluid of light in a paraxial-propagation geometry was recently reported in \cite{Vocke2015}.

While typical defects in microcavity devices are rigidly bound to the semiconductor host material \cite{Amo2009, Carusotto2013, Nardin2011, Sanvitto2011, Amo2011, Grosso2011}, the propagating geometry makes it possible to consider situations with movable and/or deformable obstacles, such as dielectric plates or rods immersed in a liquid-state nonlinear dielectric. This condition is essential to have an observable mechanical displacement and/or a deformation of the obstacle in response to the radiation pressure. According to our predictions, the transition to a superfluid state is in fact signaled by a sudden drop of the drag force corresponding to the radiation pressure and, therefore, of the optomechanical deformation of the obstacle. While the present work is focused on the classical contribution that dominates the drag force at the mean-field level, a full quantum theory of light propagation \cite{Larre2014} is needed to properly investigate the quantum drag force that was anticipated to arise from the scattering of quantum fluctuations \cite{Roberts2005}.

The article is structured as follows. In Sec.~\ref{Sec:PhysicalSystemAndTheoreticalModel}, we introduce the physical system under consideration and we review the theoretical formalism used to describe light propagation in the investigated nonlinear medium and scattering on the obstacle. The signatures of superfluidity in the intensity patterns of light are discussed in Sec.~\ref{Sec:LightSuperfluidity} in both a one-dimensional geometry with a plate-shaped obstacle and a two-dimensional geometry with a rod-shaped one. The theoretical framework to calculate the electromagnetic forces exerted by the fluid of light onto the obstacle is presented in Sec.~\ref{Sec:ElectromagneticForce}. Some quantitative predictions for the actual magnitude of the mechanical deformation that one may realistically expect in an experiment are discussed in Sec.~\ref{Sec:MechanicalDeformation}. Conclusions are finally drawn in Sec.~\ref{Sec:Conclusion}.

\section{Physical system and theoretical model}
\label{Sec:PhysicalSystemAndTheoreticalModel}

\begin{figure}
\includegraphics[width=\linewidth]{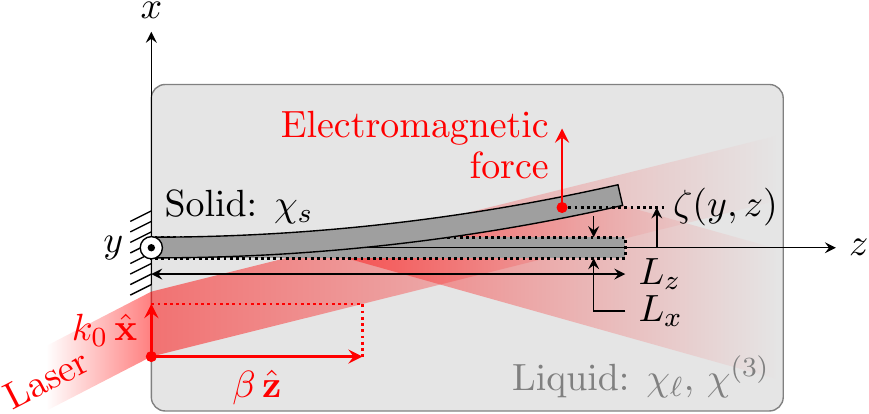}
\caption{(Color online) Sketch of the considered experimental setup, viewed from above.}
\label{ExperimentalSetup}
\end{figure}

A sketch of the physical system considered in this work is shown in Fig.~\ref{ExperimentalSetup}. A solid and transparent object of dielectric susceptibility $\chi_{s}$ is immersed into a large vessel filled with a nonlinear optical liquid of linear susceptibility $\chi_{\ell}$ and Kerr-nonlinearity coefficient $\chi^{(3)}$. Both the solid object and the liquid are devoid of free charges and nonmagnetic. The front (at $z=0$) $(x,y)$ face of the object is mechanically clamped to the tank while the rest (extending for a length $L_{z}$ in the $z$ direction) is free to move in the liquid bath. The coordinate origin corresponds to the center of the clamped face and the $y$ axis to the vertical direction.

In the following, we shall consider two geometrical shapes for the obstacle. An effectively one-dimensional dynamics for the photon fluid is obtained with a plate of thickness $L_{x}$ in the $x$ direction and very large (approximately infinite) size in the $y$ direction, so that the light-field amplitude does not depend on $y$. On the other hand, a full two-dimensional dynamics is recovered using a rod-shaped obstacle. To simplify the calculation of the electromagnetic force, we will consider a rod with a rectangular cross section of sides $L_{x}$ and $L_{y}$, typically such that $L_{y}\gg L_{x}$.

The system is illuminated by a wide monochromatic-plane-wave laser beam incident along a direction close to the $z$ axis. Within the framework of the well-known paraxial and slowly-varying-envelope approximations (see, e.g., Refs.~\cite{Rosanov2002, Agrawal2007}, but also Ref.~\cite{Carusotto2014}), we can expand the electric field $\mathbf{E}(\mathbf{x},t)=\mathrm{Re}[\boldsymbol{\mathcal{E}}(\mathbf{x})\,\mathrm{e}^{\mathrm{i}(\beta z-\omega t)}]$ [where $\mathbf{x}=(x,y,z)$] of the laser wave as the product of a slowly-varying spatial envelope $\boldsymbol{\mathcal{E}}(\mathbf{x})$ and a rapidly varying carrier of pulsation $\omega$ and wavenumber $\beta=(1+\chi_{\ell})^{1/2}\,\omega/c$ in the positive-$z$ direction, $c$ denoting the free-space speed of light.

Neglecting the polarization degrees of freedom, this yields a propagation equation for the (scalar) envelope $\mathcal{E}(\mathbf{x})$ of the electric field in a form closely analogous to the Gross--Pitaevskii equation of dilute Bose--Einstein fluids \cite{Dalfovo1999, Pitaevskii2003},
\begin{equation}
\label{GrossPitaevskiiEquation1}
\mathrm{i}\,\frac{\partial\mathcal{E}}{\partial z}=-\frac{1}{2\,\beta}\,\bigg(\frac{\partial^{2}\mathcal{E}}{\partial x^{2}}+\frac{\partial^{2}\mathcal{E}}{\partial y^{2}}\bigg)+V(\mathbf{x})\,\mathcal{E}+g\,|\mathcal{E}|^{2}\,\mathcal{E},
\end{equation}
where the longitudinal coordinate $z$ plays the role of time (in this respect, we will frequently use the adjectives ``stationary'' or ``steady'' to designate something which does not depend on $z$) and the effective photon mass equals $\beta$. In Eq.~\eqref{GrossPitaevskiiEquation1},
\begin{equation}
\label{ExternalPotential}
V(\mathbf{x})\simeq-\frac{\beta\,(\chi_{s}-\chi_{\ell})}{2\,(1+\chi_{\ell})}\,\Theta(L_{x}/2-|x|)\,\Theta(L_{y}/2-|y|),
\end{equation}
where $\Theta$ is the Heaviside step function, is the external potential arising from the refractive-index difference between the obstacle (``$s$'') and the liquid (``$\ell$''), and
\begin{equation}
\label{CouplingConstant}
g=-\frac{\beta\,\chi^{(3)}}{2\,(1+\chi_{\ell})}
\end{equation}
is the photon-photon contact-interaction constant, proportional to the Kerr coefficient $\chi^{(3)}$ of the liquid bath. In Eq.~\eqref{ExternalPotential}, we have assumed that the shape of the obstacle does not depend on $z$. This is accurate provided the deformation of the obstacle in response to the optomechanical force that it undergoes is small with respect to its transverse size (and therefore negligible at the level of the optical-field dynamics). The validity of this assumption will be checked a posteriori in Sec.~\ref{Sec:MechanicalDeformation}. In what follows, we will furthermore restrict our attention to the case of a self-defocusing Kerr nonlinearity ($\chi^{(3)}<0$), which corresponds to repulsive photon-photon interactions ($g>0$) and prevents the occurrence of dynamical instabilities in the fluid of light \cite{Carusotto2014}.

The initial condition (i.e., at $z=0$) is fixed by the transverse profile of the incident beam, that we take slightly tilted by a positive angle $\theta=\arctan(c\,k_{0}/\omega)\simeq c\,k_{0}/\omega$ away from the $z$ axis, as illustrated in Fig.~\ref{ExperimentalSetup}. This gives a small wavenumber $k_{0}>0$ to the photons in the $x$ direction,
\begin{equation}
\label{TransverseWavenumber}
\mathcal{E}(x,y,z=0)=\mathcal{E}(x,y)\,\mathrm{e}^{\mathrm{i}k_{0}x},
\end{equation}
and is similar to what has been proposed in Ref.~\cite{Pomeau1993} to study vorticity generation in a nonlinear-propagating-optics configuration. The overall envelope $\mathcal{E}(x,y)$ in Eq.~\eqref{TransverseWavenumber} is supposed to have a very wide top-hat shape, so that it can be approximately considered uniform, $\mathcal{E}(x,y)=\mathcal{E}_{0}=\mathrm{cst}$, in the region of interest. In the absence of any obstacle, the field then has a plane-wave evolution in the $z$ direction:
\begin{equation}
\label{PlaneWaveEvolution}
\mathcal{E}(x,y,z)=\mathcal{E}_{0}\,\mathrm{e}^{\mathrm{i}k_{0}x}\,\mathrm{e}^{-\mathrm{i}\varkappa z},
\end{equation}
where the wavenumber $\varkappa=k_{0}^{2}/(2\,\beta)+g\,|\mathcal{E}_{0}|^{2}$ corresponds to the chemical potential in the theory of weakly-interacting Bose gases at zero temperature \cite{Dalfovo1999, Pitaevskii2003}.

\section{Light superfluidity}
\label{Sec:LightSuperfluidity}

As a first step, we need to calculate the evolution of the transverse field during the propagation along the $z$ axis. In particular, we shall concentrate on the stationary field profiles that the incident beam of light assumes after long propagation distances. Unless otherwise specified, we shall restrict our attention to very wide incident beams moving in the positive-$x$ direction and neglect all effects stemming from the edges of the beam waist.

First, in Sec.~\ref{SubSec:OneDimensionalPlateGeometry}, we will investigate a one-dimensional configuration where superfluidity affects the nonlinear tunneling across a plate [located at $x=0$, which separates the upstream region ($x<0$) from the downstream one ($x>0$)]. Then, in Sec.~\ref{SubSec:TwoDimensionalRodGeometry}, we shall address a two-dimensional geometry where superfluidity is studied in terms of the scattering of the photon fluid on a spatially localized obstacle. From a hydrodynamic perspective, this latter configuration aims at providing an idealized, yet reasonably realistic, model of the interaction of a flowing fluid of light with the rough surface of its container.

\subsection{One-dimensional plate geometry}
\label{SubSec:OneDimensionalPlateGeometry}

In the case where $L_{y}=\infty$ (corresponding to a plate of infinite size in the $y$ direction), the optical field does not depend on $y$ and the evolution equation \eqref{GrossPitaevskiiEquation1} becomes one dimensional. In this case, analytical insight of the stationary solutions is available and the main remaining difficulty concerns how these latter can be actually reached in a realistic experiment.

\subsubsection{Stationary intensity profiles}
\label{SubSubSec:StationaryIntensityProfiles}

\begin{figure*}
\includegraphics[width=0.75\linewidth]{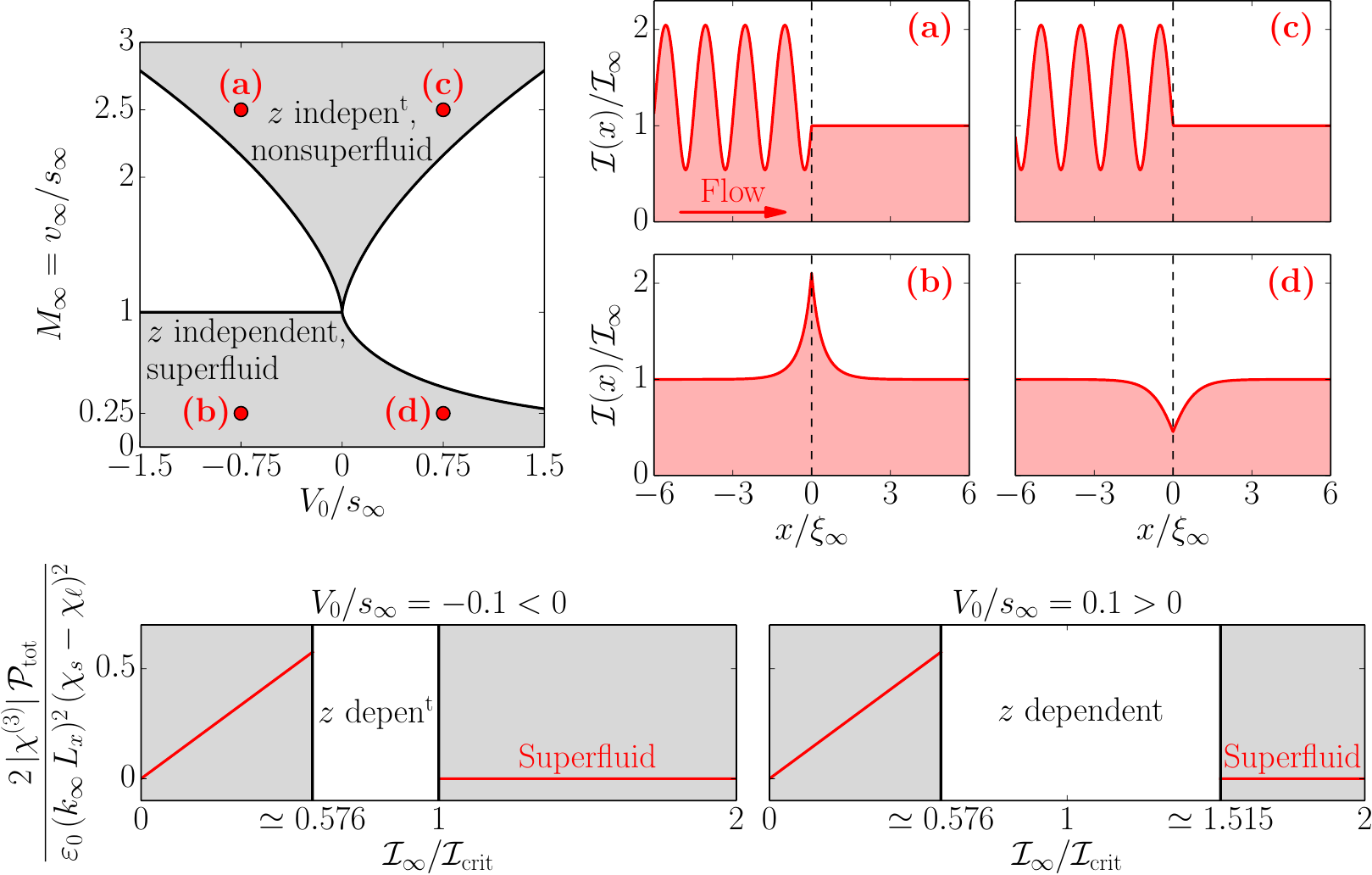}
\caption{(Color online) \textbf{One-dimensional plate geometry.---} Upper left panel: Domain of existence of the stationary ($z$-independent) solutions in the $(V_{0}/s_{\infty},M_{\infty}=v_{\infty}/s_{\infty})$ plane (gray-shaded area); the red tags indicate the parameters used for plotting the intensity patterns (a)--(d) in the upper right of the figure (the vertical dashed line indicates the position of the $\delta$-peak obstacle). Lower panels: Nonmonotonic behavior---at a fixed asymptotic photon-fluid velocity $v_{\infty}=k_{\infty}/\beta$---of the radiation pressure \eqref{TotalPolarizationPressure1}--\eqref{TotalPolarizationPressure3} as a function of the far-downstream intensity $\mathcal{I}_{\infty}$, normalized to the critical Landau intensity $\mathcal{I}_{\mathrm{crit}}=\beta\,v_{\infty}^{2}/g=k_{\infty}^{2}/(\beta\,g)$, in the two regimes $V_{0}\lessgtr0$ ($\chi_{s}\gtrless\chi_{\ell}$); in each case, the white region indicates the window in which Eq.~\eqref{GrossPitaevskiiEquation2} does not admit a stable stationary solution.}
\label{1DPlateGeometry}
\end{figure*}

In the presence of the plate ($\chi_{s}\neq\chi_{\ell}$), the scattering on the susceptibility jump $|\chi_{s}-\chi_{\ell}|$ is responsible for a complex evolution which, for suitable incident parameters, eventually tends to a stationary solution satisfying the $z$-independent equation
\begin{equation}
\label{GrossPitaevskiiEquation2}
\left(\frac{k_{\infty}^{2}}{2\,\beta}+g\,\mathcal{I}_{\infty}\right)\mathcal{E}=-\frac{1}{2\,\beta}\,\frac{\mathrm{d}^{2}\mathcal{E}}{\mathrm{d}x^{2}}+V_{0}\,\delta(x)\,\mathcal{E}+g\,|\mathcal{E}|^{2}\,\mathcal{E},
\end{equation}
with a purely-outgoing plane wave of wavenumber $k_{\infty}>0$ and constant intensity $\mathcal{I}_{\infty}=|\mathcal{E}_{\infty}|^{2}$ as boundary condition in the positive- and large-$x$ region. The $\delta$ approximation for the square potential \eqref{ExternalPotential} in Eq.~\eqref{GrossPitaevskiiEquation2}, where
\begin{equation}
\label{ExternalPotentialMagnitude}
V_{0}=-\frac{\beta\,L_{x}\,(\chi_{s}-\chi_{\ell})}{2\,(1+\chi_{\ell})},
\end{equation}
is accurate provided $L_{x}$ is much smaller than both $1/k_{\infty}$ and the asymptotic healing length $\xi_{\infty}=1/(\beta\,g\,\mathcal{I}_{\infty})^{1/2}$.

Under this approximation, analytical solutions to the nonlinear equation \eqref{GrossPitaevskiiEquation2}, as well as formulas explicitly precising their domain of existence as a function of $V_{0}$, $k_{\infty}$, and $\mathcal{I}_{\infty}$, are available in the literature \cite{Pavloff2002, Hakim1997, Leboeuf2001}. A review of these results is reported in Appendix \ref{Sec:Appendix}. As it is shown in Fig.~\ref{1DPlateGeometry}, three different regimes can be identified, depending on the sign of $V_{0}$ and on the value of the so-called Mach number
\begin{equation}
\label{MachNumber}
M_{\infty}=\frac{v_{\infty}}{s_{\infty}},
\end{equation}
where $v_{\infty}=k_{\infty}/\beta$ denotes the velocity of the fluid of light and $s_{\infty}=(g\,\mathcal{I}_{\infty}/\beta)^{1/2}=1/(\beta\,\xi_{\infty})$ the speed of sound \cite{Carusotto2014} far downstream from the obstacle ($x\gg\xi_{\infty}$). At a given asymptotic flow speed $v_{\infty}$, the Mach number $M_{\infty}$ can be written in terms of the light intensity $\mathcal{I}_{\infty}$ as $M_{\infty}=(\mathcal{I}_{\mathrm{crit}}/\mathcal{I}_{\infty})^{1/2}$, where $\mathcal{I}_{\mathrm{crit}}=\beta\,v_{\infty}^{2}/g=k_{\infty}^{2}/(\beta\,g)$ is the critical intensity for superfluidity, as defined in the so-called Landau criterion \cite{Landau1941, Carusotto2013}. Even if the link to superfluidity is rarely made in explicit terms, nonlinear-tunneling experiments similar to the one we are proposing have been recently performed by several groups \cite{Jia2007, Wan2010, Sun2012}.

For low flow speeds/high intensities (that is, for small Mach number $M_{\infty}=v_{\infty}/s_{\infty}$), the obstacle produces a localized perturbation in the intensity profile $\mathcal{I}(x)=|\mathcal{E}(x)|^{2}$ [panels (b) and (d) of Fig.~\ref{1DPlateGeometry}]; this latter exponentially recovers its unperturbed value $\mathcal{I}_{\infty}$ on both sides away from the plate. Most remarkably, the light intensity remains in this case symmetric with respect to the origin. This regime corresponds to the superfluid behavior first demonstrated in Ref.~\cite{Amo2009}.

For high flow speeds/low intensities (that is, for large Mach number $M_{\infty}=v_{\infty}/s_{\infty}$), a periodic intensity modulation due to the interference of the incident and reflected waves appears in the negative-$x$ region. In the case of a weakly perturbing obstacle ($|V_{0}|/s_{\infty}\ll1$), this modulation can be interpreted as the result of a Cherenkov radiation of Bogoliubov excitations by the obstacle (see, e.g., Ref.~\cite{Carusotto2006}). For strongly disturbing obstacles instead, it is altered by the nonlinearity and takes the form of a cnoidal wave \cite{Marburger1978, Kamchatnov2000}. Examples of such patterns are shown in the panels (a) and (c) of Fig.~\ref{1DPlateGeometry}.

In between these two regimes [corresponding to the white domain of the $(V_{0}/s_{\infty},M_{\infty})$ plane shown in the upper left panel of Fig.~\ref{1DPlateGeometry}], the flow is $z$ dependent and so can no longer be described by a stationary solution of Eq.~\eqref{GrossPitaevskiiEquation2}. In that case, a train of solitons can for instance be periodically emitted by the defect \cite{Leboeuf2001}; this regime is the one-dimensional analog of the vortex phase experimentally observed in Refs.~\cite{Nardin2011, Sanvitto2011}. When $V_{0}<0$, i.e., when $\chi_{s}>\chi_{\ell}$, superfluidity extends up to $M_{\infty}=1$ ($v_{\infty}=s_{\infty}$), i.e., up to the Landau prediction \cite{Landau1941} for Bose--Einstein condensates. When $V_{0}>0$, i.e., when $\chi_{s}<\chi_{\ell}$, superfluidity is instead lost at a lower $M_{\infty}$.

\subsubsection{Reaching the stationary state}
\label{SubSubSec:ReachingTheStationaryState}

\begin{figure}
\includegraphics[width=\linewidth]{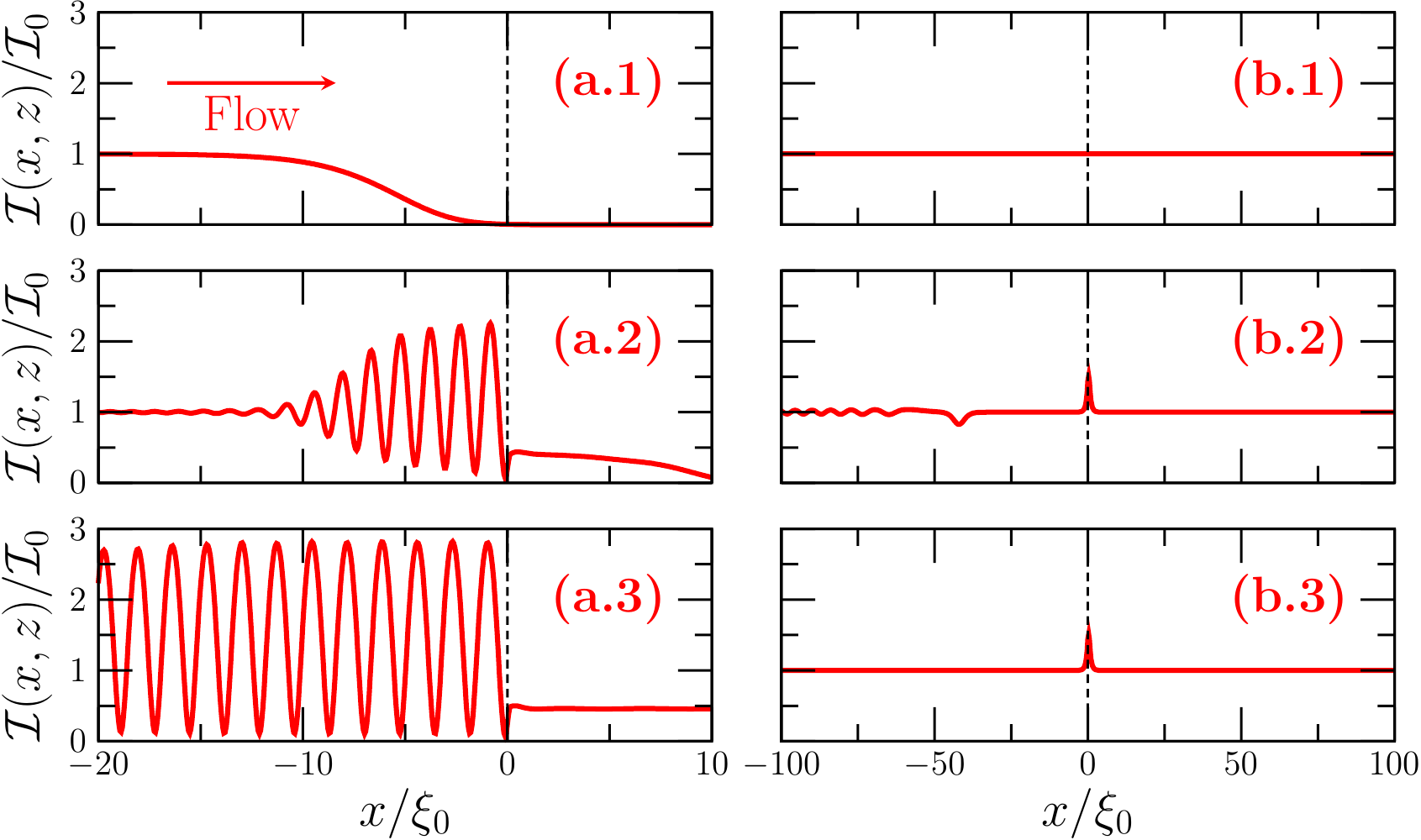}
\caption{(Color online) \textbf{One-dimensional plate geometry.---} Snapshots of the normalized intensity profile $\mathcal{I}(x,z)/\mathcal{I}_{0}$ at different propagation distances $z$, from the incident spot (a.1, b.1) towards the stationary state (a.3, b.3) showing a (nonlinear) Bogoliubov--Cherenkov modulation [column (a), $M_{0}=v_{0}/s_{0}=2.2$] or a superfluid behavior [column (b), $M_{0}=0.5$]. The vertical dashed line at the origin ($x=0$) indicates the position of the $\delta$-peak obstacle. The propagation distance $z$ is such that $k_{0}^{2}\,z/\beta=0$ (a.1, b.1), $25$ (a.2, b.2), and $125$ (a.3, b.3).}
\label{EvolutionIntensityProfiles}
\end{figure}

Even though it is in principle possible to design an incident light profile with the exact shape of the stationary state, it could be experimentally much more convenient to start with a wide intensity spot and let the steady state be spontaneously reached after some propagation distance $z$. In the one-dimensional configuration studied in this section, the choice of a suitable shape for the incident beam is a nontrivial task and must be specifically designed in the different considered cases. We present in Fig.~\ref{EvolutionIntensityProfiles} two specific examples obtained from a numerical integration of the one-dimensional version of Eq.~\eqref{GrossPitaevskiiEquation1}.

The superfluid regime (b.3), for which $v_{\infty}<s_{\infty}$, can be created by using a wide incident spot with a top-hat profile of in-plane wavenumber $k_{0}$ and peak intensity $\mathcal{I}_{0}$ [and, correspondingly, velocity $v_{0}=k_{0}/\beta$, speed of sound $s_{0}=(g\,\mathcal{I}_{0}/\beta)^{1/2}$, and healing length $\xi_{0}=1/(\beta\,s_{0})$] encompassing an attractive obstacle ($V_{0}<0$, i.e., $\chi_{s}>\chi_{\ell}$) [panel (b.1)]. In that case, after a transient emission of elementary excitations [panel (b.2)], the asymptotic parameters $k_{\infty}$ and $\mathcal{I}_{\infty}$ exactly match the incident ones.

The (nonlinear) Bogoliubov--Cherenkov regime (a.3), for which $v_{\infty}>s_{\infty}$, can be obtained by designing a wide incident spot localized upstream from the obstacle [panel (a.1)]: Scattering on this latter [panel (a.2)] automatically generates the desired stationary pattern (a.3). Note that, in order to avoid forming the transonic-interface configuration described in Ref.~\cite{Kamchatnov2012}, $k_{0}$ must be chosen sufficiently large. In contrast to the previous case, here the asymptotic momentum and intensity are not straightforwardly related to the incident ones, but can be, of course, easily measured from the light emerging from the system.

\subsection{Two-dimensional rod geometry}
\label{SubSec:TwoDimensionalRodGeometry}

\begin{figure*}
\includegraphics[width=0.75\linewidth]{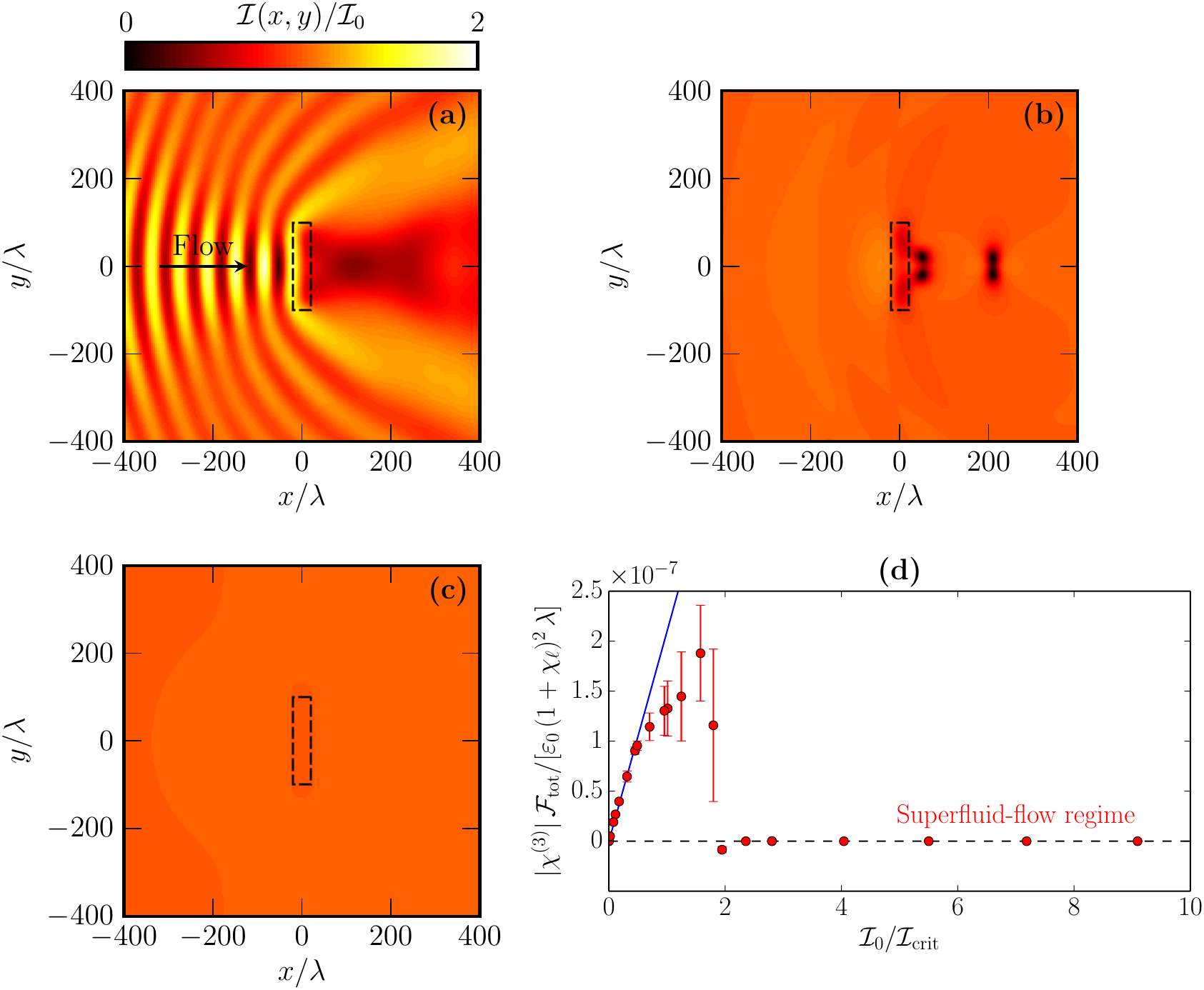}
\caption{(Color online) \textbf{Two-dimensional rod geometry.---} Panels (a)--(c): Stationary (that is, $z$-independent) light-intensity profiles $\mathcal{I}(x,y)$ (normalized to the incident intensity $\mathcal{I}_{0}$) in the deeply nonsuperfluid regime [panel (a), $M_{0}=2.98$], in the vortex-nucleation regime [panel (b), $M_{0}=0.75$], and in the superfluid regime [panel (c), $M_{0}=0.25$]; the transverse cross section of the rod-shaped obstacle is indicated by the dashed rectangle; propagation distance: $z/\lambda=9\times10^{4}$ ($\lambda=2\pi/\beta$ is the wavelength of the laser in the liquid). Panel (d): Behavior---at a fixed far-upstream velocity $v_{0}=k_{0}/\beta$---of the electromagnetic force $\mathcal{F}_{\mathrm{tot}}$ (defined in the first paragraph of Sec.~\ref{SubSec:TwoDimensionalRodGeometryForce}) as a function of the input intensity $\mathcal{I}_{0}$ normalized to the critical Landau intensity $\mathcal{I}_{\mathrm{crit}}=\beta\,v_{0}^{2}/g=k_{0}^{2}/(\beta\,g)$ (as $v_{0},k_{0}=v_{\infty},k_{\infty}$ in the two-dimensional rod configuration, this corresponds to the same normalization as the one used in the lowest panels of Fig.~\ref{1DPlateGeometry}); the points (vertical error bars) correspond to the average (standard deviation) of the force over the propagation-distance window $z/\lambda\in[4.5\times10^{4},9\times10^{4}]$ and the blue straight line indicates the linear behavior of the radiation force in the low-$\mathcal{I}_{0}$ regime. Obstacle's amplitude: $(\chi_{s}-\chi_{\ell})/(1+\chi_{\ell})=5\times10^{-5}$; obstacle's size: $L_{x}/\lambda=40$, $L_{y}/\lambda=200$.}
\label{2DRodGeometry}
\end{figure*}

The situation is in some manner simpler in the fully two-dimensional case where the obstacle has, e.g, the shape of a rod oriented in the $z$ direction. Examples of stationary-field configurations after a long $z$ propagation are displayed in panels (a)--(c) of Fig.~\ref{2DRodGeometry} for a constant incident wavenumber $k_{0}$ in the positive-$x$ direction but different values of the incident light intensity $\mathcal{I}_{0}$. Specifically, we shall focus on the case of a rod with a rectangular cross section \cite{NoteNumerics} such that $L_{y}\gg L_{x}$, which will facilitate the analysis of the electromagnetic force in the next section. As a most remarkable feature of the two-dimensional case, it is worth stressing how, in contrast to the one-dimensional case, the intensity and the speed of the fluid of light far downstream from the obstacle recover their unperturbed incident values \cite{Carusotto2014}: $\mathcal{I}_{\infty}=\mathcal{I}_{0}$ [the asymptotic sound speeds $s_{\infty}=(g\,\mathcal{I}_{\infty}/\beta)^{1/2}$ and $s_{0}=(g\,\mathcal{I}_{0}/\beta)^{1/2}$ are consequently equal] and $v_{\infty}=v_{0}$ (the asymptotic Mach numbers $M_{\infty}=v_{\infty}/s_{\infty}$ and $M_{0}=v_{0}/s_{0}$ are then identical).

As usual, a superfluid behavior is numerically found in the high-intensity regime ($\mathcal{I}_{0}\gg\mathcal{I}_{\mathrm{crit}}$), where the flow speed is lower than the speed of sound, $v_{0}\ll s_{0}$ (or, equivalently, $v_{\infty}\ll s_{\infty}$): The only effect of the obstacle is to generate a localized perturbation in the fluid profile [panel (c)]. In the opposite limit ($\mathcal{I}_{0}\ll\mathcal{I}_{\mathrm{crit}}$), that is, when $v_{0}\gg s_{0}$ (or, equivalently, $v_{\infty}\gg s_{\infty}$), Bogoliubov--Cherenkov waves upstream from the obstacle are a clear signature of a superfluidity breakdown [panel (a)]. In between, superfluidity can be broken by a different mechanism \cite{Frisch1992} due to the quasiperiodic nucleation of vortex pairs [panel (b)], as experimentally observed in planar-microcavity polariton fluids \cite{Nardin2011, Sanvitto2011} and in propagating nonlinear optics \cite{Wan2008}.

\section{Electromagnetic force}
\label{Sec:ElectromagneticForce}

The intensity profiles discussed in the previous section constitute the starting point of the calculation of the force exerted by the fluid of light on the dielectric obstacle. As this latter and the surrounding liquid are made of neutral and nonmagnetic dielectrics, we can make use of the theory of electromagnetic forces induced by the light field on the oscillating dipoles of matter. Our treatment of these forces is based on recent works by Barnett and Loudon \cite{Barnett2006, Loudon2006}. To estimate the actual deformation of the obstacle, the electromagnetic forces will then be inserted into the mechanical equations describing the static equilibrium of the full system, composed of the solid dielectric obstacle and the surrounding liquid: In addition to the direct bulk electromagnetic force, the obstacle actually also feels the mechanical pressure exerted by the dielectric liquid on its surface. Our choice of a rod-shaped geometry with a rectangular cross section aims at reducing as much as possible the complexity of the mechanical calculation.

Following Refs.~\cite{Barnett2006, Loudon2006}, the general expression of the electromagnetic force density felt by a generic dielectric is [we write down $(x_{1},x_{2},x_{3})=(x,y,z)$]
\begin{equation}
\label{PolarizationForceDensity}
\mathbf{f}(\mathbf{x},t)=\sum_{j\in\{1,2,3\}}P_{j}\,\frac{\partial\mathbf{E}}{\partial x_{j}}+\frac{\partial\mathbf{P}}{\partial t}\times\mathbf{B},
\end{equation}
where $\mathbf{E}(\mathbf{x},t)$ and $\mathbf{B}(\mathbf{x},t)$ are the electric and magnetic fields of the light wave and $\mathbf{P}(\mathbf{x},t)$ is the polarization density of the medium, including both linear and nonlinear contributions. In our case, the polarization densities of the solid obstacle and the liquid bath are respectively given by
\begin{subequations}
\label{PolarizationDensityLiquidSolid}
\begin{align}
\label{PolarizationDensitySolid}
&\mathbf{P}_{s}(\mathbf{x},t)=\varepsilon_{0}\,\chi_{s}\,\mathbf{E} \\
\label{PolarizationDensityLiquid}
\text{and}\quad&\mathbf{P}_{\ell}(\mathbf{x},t)=\varepsilon_{0}\,(\chi_{\ell}+\chi^{(3)}\,|\boldsymbol{\mathcal{E}}|^{2})\,\mathbf{E},
\end{align}
\end{subequations}
where $\varepsilon_{0}$ denotes the permittivity of free space. Taking advantage of the Maxwell--Faraday equation to express $\mathbf{B}$ as a function of $\mathbf{E}$ in Eq.~\eqref{PolarizationForceDensity}, simple algebraic manipulations \cite{Barnett2006, Loudon2006} in the case of a monochromatic light field of pulsation $\omega$---we set $\mathbf{E}(\mathbf{x},t)=\mathrm{Re}[\tilde{\mathbf{E}}(\mathbf{x})\,\mathrm{e}^{-\mathrm{i}\omega t}]$ and $\mathbf{P}(\mathbf{x},t)=\mathrm{Re}[\tilde{\mathbf{P}}(\mathbf{x})\,\mathrm{e}^{-\mathrm{i}\omega t}]$; in our work (see Sec.~\ref{Sec:PhysicalSystemAndTheoreticalModel}), one has $\tilde{\mathbf{E}}(\mathbf{x})=\boldsymbol{\mathcal{E}}(\mathbf{x})\,\mathrm{e}^{\mathrm{i}\beta z}$---lead to the following expression for the $i$th ($i\in\{1,2,3\}$) component of the time-averaged electromagnetic force density:
\begin{align}
\notag
\bar{f}_{i}(\mathbf{x})&=\int_{0}^{2\pi/\omega}f_{i}(\mathbf{x},t)\,\frac{\mathrm{d}t}{2\pi/\omega} \\
\label{PolarizationForceDensityBis}
&=\frac{1}{2}\sum_{j\in\{1,2,3\}}\mathrm{Re}\bigg(\tilde{P}_{j}^{\ast}\,\frac{\partial\tilde{E}_{j}}{\partial x_{i}}\bigg),
\end{align}
Given the symmetry of our setup with respect to the $y=0$ plane, we can focus our attention on the $x$ component of the electromagnetic force. Within the assumed paraxial-propagation regime, we can approximate the light-wave polarization to be everywhere parallel to the $y$ axis \cite{NotePolarization}.

Integrating Eq.~\eqref{PolarizationForceDensityBis} over a thin volume of the obstacle of transverse sizes $\mathrm{d}y$ and $\mathrm{d}z$ encompassing its thickness in the $x$ direction, we get to an electromagnetic pressure acting on the solid obstacle at position $(y,z)$ given by
\begin{align}
\notag
\mathcal{P}_{s}(y,z)&=\frac{1}{2}\int_{-L_{x}/2}^{L_{x}/2}\mathrm{Re}\bigg[\tilde{P}_{s}^{\ast}(\mathbf{x})\,\frac{\partial\tilde{E}}{\partial x}(\mathbf{x})\bigg]\,\mathrm{d}x \\
\notag
&=\frac{\varepsilon_{0}\,\chi_{s}}{4}\int_{-L_{x}/2}^{L_{x}/2}\frac{\partial}{\partial x}\,|\tilde{E}(\mathbf{x})|^{2}\,\mathrm{d}x \\
\label{PolarizationPressureSolid}
&=\frac{\varepsilon_{0}\,\chi_{s}}{4}\,\Big[\mathcal{I}(L_{x}/2,y,z)-\mathcal{I}(-L_{x}/2,y,z)\Big],
\end{align}
where $\mathcal{I}(\mathbf{x})=|\tilde{E}(\mathbf{x})|^{2}=|\mathcal{E}(\mathbf{x})|^{2}$ is the electric-field intensity.

Of course, a similar electromagnetic force acts also on the surrounding liquid. Assuming the liquid to be incompressible, this electromagnetic force only results in a spatial variation of the local liquid pressure $\Pi_{\ell}(\mathbf{x})$ according to the hydrostatic-equilibrium condition
\begin{equation}
\label{HydrostaticEquilibrium}
\bar{\mathbf{f}}_{\ell}(\mathbf{x})=\nabla\Pi_{\ell}(\mathbf{x}).
\end{equation}
Assuming that the liquid pressure recovers the atmospheric pressure in the dark region outside the laser field and reminding that $\chi_{\ell}$ and $\chi^{(3)}$ are assumed to be spatially homogeneous, the liquid-pressure difference between the two interfaces solid/liquid parallel to the $(y,z)$ plane at $(x=\mp\,L_{x}/2,y,z)$ reads
\begin{align}
\notag
\Delta\Pi_{\ell}(y,z)&\left.=\Pi_{\ell}(-L_{x}/2,y,z)-\Pi_{\ell}(L_{x}/2,y,z)\right. \\
\notag
&\left.=\frac{\varepsilon_{0}}{4}\,\bigg[\chi_{\ell}+\frac{\chi^{(3)}}{2}\,\mathcal{I}(-L_{x}/2,y,z)\bigg]\right. \\
\label{PolarizationPressureLiquid}
&\left.\phantom{=}\times\mathcal{I}(-L_{x}/2,y,z)-(L_{x}\longleftrightarrow-L_{x}).\right.
\end{align}

Since the obstacle is subject to the direct bulk electromagnetic force \eqref{PolarizationPressureSolid} and the indirect liquid-pressure effect \eqref{PolarizationPressureLiquid}, the total force per unit area acting on the obstacle can be written in terms of the light intensity profile $\mathcal{I}(\pm\,L_{x}/2,y,z)$ in the following way:
\begin{align}
\notag
\mathcal{P}_{\mathrm{tot}}(y,z)&\left.=\mathcal{P}_{s}(y,z)+\Delta\Pi_{\ell}(y,z)\right. \\
\notag
&\left.=\frac{\varepsilon_{0}}{4}\,\bigg[\chi_{s}-\chi_{\ell}-\frac{\chi^{(3)}}{2}\,\mathcal{I}(L_{x}/2,y,z)\bigg]\right. \\
\label{TotalPolarizationPressure}
&\left.\phantom{=}\times\mathcal{I}(L_{x}/2,y,z)-(L_{x}\longleftrightarrow-L_{x}).\right.
\end{align}
This quantity can straightforwardly be extracted from the calculations exposed in the previous section.

\subsection{One-dimensional plate geometry}
\label{SudSec:OneDimensionalPlateGeometryForce}

In a one-dimensional thin-plate geometry well in the stationary state, we can neglect the $(y,z)$ dependence and further simplify the expression \eqref{TotalPolarizationPressure} making use of the estimate
\begin{equation}
\label{DL}
\mathcal{I}(L_{x}/2)-\mathcal{I}(-L_{x}/2)\simeq\frac{L_{x}}{2}\,\bigg[\frac{\mathrm{d}\mathcal{I}}{\mathrm{d}x}(0^{+})+\frac{\mathrm{d}\mathcal{I}}{\mathrm{d}x}(0^{-})\bigg],
\end{equation}
valid in the thin-plate approximation ($L_{x}\ll\xi_{\infty}$). Using the well-known analytical solutions of Eq.~\eqref{GrossPitaevskiiEquation2} (see Appendix \ref{Sec:Appendix} for details) and neglecting the (small) nonlinear correction, the resulting force per unit surface $\mathcal{P}_{\rm tot}$ experienced by the plate is, in the nonsuperfluid regime,
\begin{equation}
\label{TotalPolarizationPressure1}
\mathcal{P}_{\mathrm{tot}}=\frac{\varepsilon_{0}\,(\beta\,L_{x})^{2}\,(\chi_{s}-\chi_{\ell})^{2}}{4\,(1+\chi_{\ell})}\,\mathcal{I}_{\infty}.
\end{equation}
On the other hand, in the superfluid regime, $\mathcal{I}(x)$ is symmetric with respect to $x=0$ and, consequently, the force is identically zero,
\begin{equation}
\label{TotalPolarizationPressure3}
\mathcal{P}_{\mathrm{tot}}=0.
\end{equation}
This behavior of the electromagnetic pressure extends to the one-dimensional strong-obstacle case the perturbative predictions of Ref.~\cite{Astrakharchik2004}. It was also fully established in Ref.~\cite{Pavloff2002} for an obstacle of arbitrary amplitude on the basis of the calculation of the stress tensor of the nonlinear fluid. In the paraxial-propagation regime considered in this work, it is also possible to get the result \eqref{TotalPolarizationPressure1}--\eqref{TotalPolarizationPressure3} for the $z$-independent electromagnetic pressure experienced by the plate from the stress tensor of the stationary Gross--Pitaevskii-like wave equation \eqref{GrossPitaevskiiEquation2} \cite{NoteStressTensor}. However, such a procedure is less direct than the one based on the physical expression \eqref{PolarizationForceDensity}--\eqref{PolarizationForceDensityBis} of the radiation force density and that is why we considered the latter to calculate the electromagnetic pressure felt by the obstacle.

This physics is summarized in the lowest panels of Fig.~\ref{1DPlateGeometry}, where the radiation pressure \eqref{TotalPolarizationPressure1}--\eqref{TotalPolarizationPressure3} is plotted as a function of the far-downstream light intensity $\mathcal{I}_{\infty}$---at a fixed value of the asymptotic photon-flow velocity $v_{\infty}=k_{\infty}/\beta$. In the low-$\mathcal{I}_{\infty}$ regime, the electromagnetic force linearly grows with the intensity, as expected for a standard linear optical process. It is interesting to note that our expression for the force per unit area recovers in the $\chi_{\ell}\to0$ and $\chi^{(3)}\to0$ limits the elementary result for the radiation pressure felt by a solid dielectric slab immersed in vacuum. In the intermediate $\mathcal{I}_{\infty}$ window (in white), there is no dynamically-stable stationary solution. As the electromagnetic field does not tend to a steady state in this region of the flow parameters and as the force is strongly sensitive to the initial conditions, one chose not to plot it. For large $\mathcal{I}_{\infty}$, one finds instead the remarkable result that the electromagnetic pressure completely vanishes: Superfluidity hinders any reflection and the fluid of light is able to freely tunnel across the plate-shaped obstacle in a frictionless way, that is, without exerting any mechanical force on it.

\subsection{Two-dimensional rod geometry}
\label{SubSec:TwoDimensionalRodGeometryForce}

In the two-dimensional configuration, there are not simple analytical techniques available (except linear-response theory but this latter is limited by the fact that the susceptibility jump $|\chi_{s}-\chi_{\ell}|$ has to be small) and we have to rely on a numerical integration of the Gross--Pitaevskii equation \eqref{GrossPitaevskiiEquation1}. The force per unit $z$-length $\mathcal{F}_{\mathrm{tot}}$ that is plotted in Fig.~\ref{2DRodGeometry} (d) is evaluated by inserting into the expression \eqref{TotalPolarizationPressure} numerically calculated intensity profiles like those shown in Fig.~\ref{2DRodGeometry} (a)--(c) and then by integrating over $y$. For the sake of simplicity, we focus our attention on the behavior of the fluid after long propagation distances $z$, i.e., when all transients due to the entrance in the nonlinear medium have gone away \cite{Carusotto2014}. In the plot of the force $\mathcal{F}_{\mathrm{tot}}$ shown in Fig.~\ref{2DRodGeometry} (d), the flow velocity $v_{0}=k_{0}/\beta$ is kept constant while the input light intensity $\mathcal{I}_{0}$ is varied.

As expected, in the low-$\mathcal{I}_{0}$ regime, ${\mathcal{F}}_{\mathrm{tot}}$ grows linearly with the light intensity, while it vanishes in the high-$\mathcal{I}_{0}$ regime: Despite the idealized rod-like shape of the obstacle, it turns out that our conclusion is fully general and the dramatic suppression of the mechanical force appears to be a generic signature of a frictionless flow of the superfluid of light around a solid defect. Note that the intensity reduction at the defect position can be very significant even in a superfluid regime, which supports the physical interpretation of light superfluidity in terms of a reduced friction by the container walls.

In between these two limiting intensity regimes, there is an intermediate window where the fluid does not get to a stationaty state, but shows a complicated $z$-dependent evolution with a quasiperiodic nucleation of vortex pairs. The error bars in Fig.~\ref{2DRodGeometry} (d) indicate the oscillation range of the ($z$-dependent) force that we have found in the numerics. As a consequence of the relatively-large obstacle strength and thickness away from the perturbative regime, nucleation of vortices appears in our simulations to extend down to low Mach numbers, i.e., for intensities $\mathcal{I}_{0}$ well above the Landau critical intensity $\mathcal{I}_{\mathrm{crit}}$ \cite{Frisch1992}.

To conclude, it is worth stressing that all these conclusions are based on a mean-field description of the flowing fluid of light in terms of a classical Gross--Pitaevskii-like wave equation which is expected to give accurate predictions in standard nonlinear media well in the weakly interacting regime. Corrections to this picture due to quantum fluctuations, as anticipated in Ref.~\cite{Roberts2005}, will be the subject of a future work.

\section{Mechanical deformation}
\label{Sec:MechanicalDeformation}

A direct way to experimentally measure the electromagnetic force acting on the obstacle is to look at the resulting deformation, which, in our configuration, consists in a bending of the material in the $x$ direction, as sketched in Fig.~\ref{ExperimentalSetup}. The magnitude of the displacement $\zeta(y,z)$ may be straightforwardly calculated using the theory of elasticity. For the sake of simplicity, we shall restrict to the case of a laterally wide rod in the $y$ direction, so that we can neglect the effect of the edges and approximate the system as infinite along the $y$ axis and subject to an electromagnetic pressure $\overline{\mathcal{P}}_{\mathrm{tot}}(z)$ uniformly distributed in this direction; moreover, since $L_{y}$ corresponds to the height of the plate, supposing $L_{y}$ large makes it possible to neglect the effect of the $y$-directed weight and buoyancy forces on the plate deformation. Under this approximation, a one-dimensional treatment of elasticity is possible in terms of a $y$-independent displacement field $\zeta$ which, assuming $|\zeta|\lesssim L_{x}\ll L_{z}$, satisfies \cite{Landau1986}
\begin{equation}
\label{EquilibriumEquation}
D\,\frac{\mathrm{d}^{4}\zeta}{\mathrm{d}z^{4}}(z)=\overline{\mathcal{P}}_{\mathrm{tot}}(z),
\end{equation}
where $D=E\,L_{x}^{3}/[12\,(1-\sigma^{2})]$ is the so-called flexural rigidity, written in terms of the Young modulus $E$ and of the Poisson ratio $\sigma$ of the material. The differential equation \eqref{EquilibriumEquation} has to be supplemented by the boundary conditions $\zeta=\partial_{z}\zeta=0$ at $z=0$ and $\partial_{z,z}\zeta=\partial_{z,z,z}\zeta=0$ at $z=L_{z}$ \cite{Landau1986}. Restricting our attention to the case of a stationary (i.e., $z$-independent) pressure $\overline{\mathcal{P}}_{\mathrm{tot}}$, we obtain the following expression for the displacement $\zeta(z)$:
\begin{equation}
\label{DisplacementField}
\zeta(z)=\frac{L_{z}^{4}}{2\,D}\left(\frac{1}{12}\,\frac{z^{4}}{L_{z}^{4}}-\frac{1}{3}\,\frac{z^{3}}{L_{z}^{3}}+\frac{1}{2}\,\frac{z^{2}}{L_{z}^{2}}\right)\overline{\mathcal{P}}_{\mathrm{tot}}.
\end{equation}

To assess the experimental feasibility of our proposal, it is essential to estimate the order of magnitude of the displacement that can be obtained for realistic parameters. Within the two-dimensional configuration of Fig.~\ref{2DRodGeometry}, one notices that the intensity right upstream from the obstacle is much larger than the downstream one and that it is just a few times larger than the incident intensity $\mathcal{I}_{0}$. As a consequence, a semiquantitative estimate of the force is straightforwardly obtained by replacing $\mathcal{I}(-L_{x}/2,y,z)$ and $\mathcal{I}(L_{x}/2,y,z)$ respectively with $\mathcal{I}_{0}$ and zero in Eq.~\eqref{TotalPolarizationPressure}.

In the specific case of a solution of ethanol doped with iodine as nonlinear optical liquid, an optical intensity in the $1~\mathrm{kW}/\mathrm{cm}^{2}$ range is required to have a nonlinear refractive-index shift $\simeq5.5\times10^{-5}$. Considering the obstacle made of fused silica, the quite large refractive-index contrast to the surrounding liquid, $\chi_{s}-\chi_{\ell}\simeq0.3$ \cite{NoteRefractiveIndexContrast}, makes the total pressure \eqref{TotalPolarizationPressure} experienced by the obstacle to be of the order of $1~\mathrm{nN}/\mathrm{mm}^{2}$. Using the mechanical constants of fused silica at room temperature, this corresponds to a deformation $\zeta(L_{z})$ of the order of a micron fraction for a $L_{x}=1~\mu\mathrm{m}$ thick and $L_{z}=1~\mathrm{mm}$ long obstacle. The fact that such a value is smaller than the obstacle thickness $L_{x}$ justifies a posteriori our hypothesis of a $z$-independent obstacle. On the other hand, such a value is well within the sensitivity range of state-of-the-art small-displacement measurements (see, e.g., \cite{Aspelmeyer2013} and references therein).

The disappearance of the deformation when entering the superfluid regime will provide a clear signature of the frictionless flow of the fluid of light, i.e., from a purely optical standpoint, of the suppressed reflection on the plate. Of course, observation of superfluidity in a medium with a lower nonlinear susceptibility $\chi^{(3)}$ would require a larger optical power, but the mechanical force would be correspondingly higher. On the other hand, the deformation effect would be strongly suppressed if a nonlinear medium in the solid instead than the liquid state was used: In this case, the mechanical rigidity of the host matrix would in fact add up to the one of the obstacle.

\section{Conclusion}
\label{Sec:Conclusion}

In this paper, we have proposed an experiment that could demonstrate the occurrence of a frictionless flow of superfluid light through and/or around a solid dielectric obstacle. For an obstacle that is not bound to a solid-state matrix, the electromagnetic force induced by the laser light results into the mechanical deformation of the obstacle. The behavior of this optical analog of the drag force as a function of the light intensity at a given flow speed (i.e., at a given angle of incidence) is strongly nonmonotonic: At low intensities, light is partially scattered by the obstacle and the deformation grows from zero linearly with the intensity; at high intensities, the deformation completely disappears, indicating a superfluid flow of photons through the obstacle (in a one-dimensional geometry with a plate-shaped obstacle) or around it (in a two-dimensional configuration with a rod-shaped obstacle). Using realistic parameters, we have checked that the actual strength of the deformation falls within the capability of state-of-the-art small-displacement measurements. An experiment along these lines would provide a crucial contribution to the understanding of the hydrodynamic properties of fluids of light, demonstrating that, also in the optical case, superfluidity is indeed associated to a drop in the drag force exerted by the fluid on obstacles stymying its flow.

\begin{acknowledgments}

We are grateful to Giuseppe C. La Rocca for his valuable input on the radiative forces and acknowledge Fernando R. Manzano and Daniele Faccio for stimulating discussions on experimental issues.

This work was supported by the ERC through the QGBE grant and by the Autonomous Province of Trento, partly through the project ``On silicon chip quantum optics for quantum computing and secure communications'' (``SiQuro'').

\end{acknowledgments}

\appendix

\section{Analytical solutions of Eq.~\texorpdfstring{\eqref{GrossPitaevskiiEquation2}}{Lg}}
\label{Sec:Appendix}

\subsection{Hydrodynamic formulation}
\label{SubSec:HydrodynamicFormulation}

The Madelung representation, which consists in writing the unknown of the stationary (i.e., $z$-independent) one-dimensional Gross--Pitaevskii equation \eqref{GrossPitaevskiiEquation2} as
\begin{equation}
\label{MadelungRepresentation}
\mathcal{E}(x)=\sqrt{\mathcal{I}(x)}\,\mathrm{e}^{\mathrm{i}\varphi(x)},
\end{equation}
makes it possible to rewrite Eq.~\eqref{GrossPitaevskiiEquation2} under the form of a system of coupled hydrodynamic-like equations verified by the laser-beam intensity $\mathcal{I}(x)=|\mathcal{E}(x)|^{2}$ and the velocity potential $\varphi(x)/\beta=\int^{x}v(x')\,\mathrm{d}x'$ [where $v(x)$ is the local speed of the light flow], namely,
\begin{subequations}
\label{HydrodynamicEquations}
\begin{align}
\label{HydrodynamicEquation1}
&\left.\frac{\mathrm{d}}{\mathrm{d}x}\bigg(\mathcal{I}\,\frac{\mathrm{d}\varphi}{\mathrm{d}x}\bigg)=0\quad\text{and}\right. \\
\notag
&\left.\frac{M_{\infty}^{2}}{2}+1=-\frac{\xi_{\infty}^{2}}{4}\,\frac{1}{\mathcal{I}}\,\frac{\mathrm{d}^{2}\mathcal{I}}{\mathrm{d}x^{2}}+\frac{\xi_{\infty}^{2}}{8}\,\frac{1}{\mathcal{I}^{2}}\,\bigg(\frac{\mathrm{d}\mathcal{I}}{\mathrm{d}x}\bigg)^{2}+\frac{\mathcal{I}}{\mathcal{I}_{\infty}}\right. \\
\label{HydrodynamicEquation2}
&\left.\phantom{\frac{M_{\infty}^{2}}{2}+1=}+\frac{\xi_{\infty}^{2}}{2}\,\bigg(\frac{\mathrm{d}\varphi}{\mathrm{d}x}\bigg)^{2}+\frac{V_{0}}{g\,\mathcal{I}_{\infty}}\,\delta(x),\right.
\end{align}
\end{subequations}
where $M_{\infty}=v_{\infty}/s_{\infty}=k_{\infty}\,\xi_{\infty}$ is the Mach number of the light flow far downstream ($x\gg\xi_{\infty}$) from the $\delta$-peak obstacle. In this region, $\mathcal{I}(x)\simeq\mathcal{I}_{\infty}$ and $\varphi(x)\simeq k_{\infty}\,x$, in such a way that, according to Eq.~\eqref{HydrodynamicEquation1}, $\partial_{x}\varphi(x)=k_{\infty}\,\mathcal{I}_{\infty}/\mathcal{I}(x)$, which, substituted into Eq.~\eqref{HydrodynamicEquation2}, yields
\begin{align}
\notag
&\left.\frac{\xi_{\infty}^{2}}{4}\,\frac{1}{\mathcal{I}}\,\frac{\mathrm{d}^{2}\mathcal{I}}{\mathrm{d}x^{2}}-\frac{\xi_{\infty}^{2}}{8}\,\frac{1}{\mathcal{I}^{2}}\,\bigg(\frac{\mathrm{d}\mathcal{I}}{\mathrm{d}x}\bigg)^{2}+\frac{M_{\infty}^{2}}{2}\,\bigg(1-\frac{\mathcal{I}_{\infty}^{2}}{\mathcal{I}^{2}}\bigg)\right. \\
\label{HydrodynamicEquationIntensity}
&+1-\frac{\mathcal{I}}{\mathcal{I}_{\infty}}=\frac{V_{0}}{g\,\mathcal{I}_{\infty}}\,\delta(x)=0\quad\textrm{for all}\quad x\neq0,
\end{align}
from which one gets, after integrating over a small-length interval containing the origin,
\begin{equation}
\label{DiscontinuityDerivative}
\frac{\mathrm{d}\mathcal{I}}{\mathrm{d}x}(0^{+})-\frac{\mathrm{d}\mathcal{I}}{\mathrm{d}x}(0^{-})=\frac{4\,V_{0}}{s_{\infty}\,\xi_{\infty}}\,\mathcal{I}(0).
\end{equation}
The problem defined by Eqs.~\eqref{HydrodynamicEquationIntensity} and \eqref{DiscontinuityDerivative} admits different solutions depending on the value of $M_{\infty}\lessgtr1$ and on the sign of $V_{0}=-\beta\,L_{x}\,(\chi_{s}-\chi_{\ell})/[2\,(1+\chi_{\ell})]$.

\subsection{Solution for \texorpdfstring{$M_{\infty}<1$}{Lg} and \texorpdfstring{$V_{0}<0$}{Lg}}
\label{SubSec:Solution1}

When $M_{\infty}<1$ and $V_{0}<0$ ($\chi_{s}>\chi_{\ell}$), the intensity profile of the laser beam is given by
\begin{align}
\notag
\frac{\mathcal{I}(x\gtrless0)}{\mathcal{I}_{\infty}}=&\left.M_{\infty}^{2}+(1-M_{\infty}^{2})\right. \\
\label{SolutionSubNeg}
&\left.\times\tanh^{-2}\bigg(\sqrt{1-M_{\infty}^{2}}\;\frac{x\pm x_{0}}{\xi_{\infty}}\bigg),\right.
\end{align}
where $x_{0}>0$ is determined by means of the matching condition \eqref{DiscontinuityDerivative}, which reads here
\begin{equation}
\label{Determinationx0SubNeg}
\frac{V_{0}}{s_{\infty}}=\frac{(1-M_{\infty}^{2})^{3/2}\coth(\sqrt{1-M_{\infty}^{2}}\,x_{0}/\xi_{\infty})}{M_{\infty}^{2}-\cosh^{2}(\sqrt{1-M_{\infty}^{2}}\,x_{0}/\xi_{\infty})}.
\end{equation}
For any $V_{0}/s_{\infty}<0$, there exists a solution $x_{0}$ (and only one) to Eq.~\eqref{Determinationx0SubNeg}, which means that, at a given Mach number $M_{\infty}<1$, one can always find an intensity profile [given by Eq.~\eqref{SolutionSubNeg}] verifying both Eq.~\eqref{HydrodynamicEquationIntensity} and Eq.~\eqref{DiscontinuityDerivative}, whatever the value of $V_{0}/s_{\infty}<0$ \cite{Leboeuf2001, Pavloff2002}.

\subsection{Solution for \texorpdfstring{$M_{\infty}<1$}{Lg} and \texorpdfstring{$V_{0}>0$}{Lg}}
\label{SubSec:Solution2}

When $M_{\infty}<1$ and $V_{0}>0$ ($\chi_{s}<\chi_{\ell}$), one gets
\begin{align}
\notag
\frac{\mathcal{I}(x\gtrless0)}{\mathcal{I}_{\infty}}=&\left.M_{\infty}^{2}+(1-M_{\infty}^{2})\right. \\
\label{SolutionSubPos}
&\left.\times\tanh^{2}\bigg(\sqrt{1-M_{\infty}^{2}}\;\frac{x\pm x_{0}}{\xi_{\infty}}\bigg),\right.
\end{align}
where $x_{0}>0$ is once more deduced from Eq.~\eqref{DiscontinuityDerivative}, i.e., here, from
\begin{equation}
\label{Determinationx0SubPos}
\frac{V_{0}}{s_{\infty}}=\frac{(1-M_{\infty}^{2})^{3/2}\tanh(\sqrt{1-M_{\infty}^{2}}\,x_{0}/\xi_{\infty})}{M_{\infty}^{2}+\sinh^{2}(\sqrt{1-M_{\infty}^{2}}\,x_{0}/\xi_{\infty})}.
\end{equation}
Equation \eqref{Determinationx0SubPos} admits two distinct solutions $x_{0}^{-}$ and $x_{0}^{+}$ (with $x_{0}^{-}<x_{0}^{+}$), and so, there exists a priori two possible intensity patterns \eqref{SolutionSubPos}, provided that \cite{Hakim1997, Leboeuf2001, Pavloff2002}
\begin{equation}
\label{BoundarySubPos}
\frac{V_{0}}{s_{\infty}}<\frac{2\,\sqrt{2}\,(1-M_{\infty}^{2})\,\sqrt{\sqrt{8\,M_{\infty}^{2}+1}-2\,M_{\infty}^{2}-1}}{\sqrt{8\,M_{\infty}^{2}+1}+4\,M_{\infty}^{2}-1}.
\end{equation}
According to Ref.~\cite{Hakim1997}, the $x_{0}\in\{x_{0}^{-},x_{0}^{+}\}$ which gives rise to a stable flow of light is $x_{0}=x_{0}^{+}$. The inequality \eqref{BoundarySubPos} defines, at a given $M_{\infty}<1$, the maximum value that $V_{0}/s_{\infty}>0$ can reach so that \eqref{SolutionSubPos} still is a solution of the problem \eqref{HydrodynamicEquationIntensity}--\eqref{DiscontinuityDerivative}.

\subsection{Solution for \texorpdfstring{$M_{\infty}>1$}{Lg} and \texorpdfstring{$V_{0}\lessgtr0$}{Lg}}
\label{SubSec:Solution3}

In the case where $M_{\infty}>1$, the radiation condition (see, e.g., Ref.~\cite{Lamb1997}) imposes that $\mathcal{I}(x)$ identically equals $\mathcal{I}_{\infty}$ in the positive-$x$ region, which implies that $\partial_{x}\mathcal{I}(0^{+})=0$ and, according to the matching equation \eqref{DiscontinuityDerivative}, that
\begin{align}
\label{DiscontinuityDerivativeSup}
\frac{\mathrm{d}\mathcal{I}}{\mathrm{d}x}(0^{-})&=-\frac{4\,V_{0}}{s_{\infty}\,\xi_{\infty}}\,\mathcal{I}_{\infty} \\
\notag
&\gtrless0\quad\textrm{if}\quad V_{0}\lessgtr0~(\chi_{s}\gtrless\chi_{\ell}).
\end{align}
In the negative-$x$ region, an infinite-range cnoidal wave is generated, the intensity profile of which can be expressed as (see, e.g., Ref.~\cite{Kamchatnov2000})
\begin{align}
\notag
\frac{\mathcal{I}(x)}{\mathcal{I}_{\infty}}=&\left.\nu_{1}+(\nu_{2}-\nu_{1})\right. \\
\label{SolutionSup}
&\left.\times\,\mathrm{sn}^{2}\bigg(\sqrt{\nu_{3}-\nu_{1}}\;\frac{x-x_{0}}{\xi_{\infty}}~\bigg|~m\bigg),\right.
\end{align}
where $\mathrm{sn}(\cdot|\cdot)$ is the Jacobi sine elliptic function, the parameter $x_{0}>0$ is determined so that $\mathcal{I}(x)$ satisfies Eq.~\eqref{DiscontinuityDerivativeSup}, $\nu_{1}$, $\nu_{2}$, and $\nu_{3}$ (with $\nu_{1}<\nu_{2}<\nu_{3}$) are the real solutions of the third-order polynomial equation \cite{NoteEquationNu}
\begin{equation}
\label{EquationNu}
\nu^{3}-(M_{\infty}^{2}+2)\,\nu^{2}+\left(2\,M_{\infty}^{2}+\frac{4\,V_{0}^{2}}{s_{\infty}^{2}}+1\right)\nu-M_{\infty}^{2}=0,
\end{equation}
and $m=(\nu_{2}-\nu_{1})/(\nu_{3}-\nu_{1})$. The $\nu_{i}$'s ($i\in\{1,2,3\}$) are real, and so, the solution \eqref{SolutionSup} exists, as long as the discriminant of Eq.~\eqref{EquationNu} stays positive, which yields, at a given $M_{\infty}>1$, the following constraint on $V_{0}/s_{\infty}$ \cite{Leboeuf2001, Pavloff2002, Larre2012}:
\begin{equation}
\label{BoundarySup}
\left|\frac{V_{0}}{s_{\infty}}\right|<\frac{\sqrt{M_{\infty}\,(M_{\infty}^{2}+8)^{3/2}+M_{\infty}^{4}-20\,M_{\infty}^{2}-8}}{4\,\sqrt{2}}.
\end{equation}


\begin{thebibliography}{99}
\bibitem{NoteNumerics}
To facilitate the numerical integration of the partial differential equation \eqref{GrossPitaevskiiEquation1}, the discontinuity of the whole electric susceptibility and of the whole nonlinear coupling constant at the edges has been smoothened out according to an exponential law.
\bibitem{NotePolarization}
Note how this choice of polarization guarantees that the electric field is continuous at the interfaces solid/liquid parallel to the $(y,z)$ plane. Even though discontinuities of the optical field at the interfaces solid/liquid parallel to the $(x,z)$ plane go beyond the paraxial and slowly-varying-envelope approximations, they do not contribute to the $x$ component of the electromagnetic force.
\bibitem{NoteStressTensor}
In the $z$-independent one-dimensional plate geometry, the electromagnetic pressure experienced by the plate may be expressed as $\mathcal{P}_{\mathrm{tot}}=\sigma(x<0)-\sigma(x>0)$, where $\sigma(x)$ denotes the stress tensor of the flowing fluid of light described by the stationary equation \eqref{GrossPitaevskiiEquation2}. It may be constructed from the Lagrangian density $\mathcal{L}(x)=\mathcal{N}\,\{-|\partial_{x}\mathcal{E}|^{2}/(2\,\beta)-[V_{0}\,\delta(x)-\varkappa]\,|\mathcal{E}|^{2}-g\,|\mathcal{E}|^{4}/2\}$ [where $\mathcal{N}=\beta/(2\,\mu_{0}\,\omega^{2})$, $\mu_{0}$ is the vacuum permeability, and $\varkappa=k_{\infty}^{2}/(2\,\beta)+g\,\mathcal{I}_{\infty}$], from which \eqref{GrossPitaevskiiEquation2} can be derived \cite{Larre2014}, as $\sigma(x)=-\partial\mathcal{L}/\partial(\partial_{x}\mathcal{E})\,\partial_{x}\mathcal{E}-\partial\mathcal{L}/\partial(\partial_{x}\mathcal{E}^{\ast})\,\partial_{x}\mathcal{E}^{\ast}+\mathcal{L}$; see, e.g., S. Weinberg, \textit{The Quantum Theory of Fields}, Vol.~1 (Cambridge University Press, Cambridge, 1995).
\end{thebibliography}
\end{document}